\documentclass[12pt]{iopart}
\usepackage{graphicx}
\usepackage{setstack,cite}
\begin{document}
\date{}

\title[Nonlocal connections]
{A nonlocal connection between certain linear and nonlinear ordinary
differential equations/oscillators}

\author{V. K. Chandrasekar$^{\dag}$, M. Senthilvelan$^{\dag}$, 
Anjan Kundu$^{\dag\dag}$ and $\;\;\;$ M. Lakshmanan$^{\dag}$}
\address{$^{\dag}$Centre for Nonlinear Dynamics, Department of Physics, 
Bharathidasan University, Tiruchirappalli - 620 024, India }
\address{$^{\dag\dag}$Saha Institute of Nuclear Physics,  
 Sector 1, Block AF, Bidhan Nagar, Kolkata - 700 064, India}

\begin{abstract}
We explore a nonlocal connection between certain linear and nonlinear ordinary
differential equations (ODEs), representing physically important oscillator
systems, and identify a class of integrable nonlinear ODEs of any
order. We also devise a method to derive explicit general solutions of the
nonlinear ODEs. Interestingly, many well known integrable models can be
accommodated into our scheme and our procedure thereby provides further 
understanding of these models.
\end{abstract}


\section{Introduction}
In a recent paper, we have shown that the modified Emden type
equation with additional linear forcing,
\begin{eqnarray}            
\ddot{x}+3kx\dot{x}+k^2x^3+\lambda x=0,\label {lam101}
\end{eqnarray}
where over dot denotes differentiation with respect to $t$ and $k$ and $\lambda$
are arbitrary parameters, exhibits certain unusual nonlinear dynamical properties
\cite{Chand1}.
For a particular sign of the control parameter, namely, $\lambda >0$, the
frequency of oscillations of the nonlinear oscillator (\ref{lam101}) 
is completely independent of amplitude and remains the
same as that of the linear harmonic oscillator, thereby showing that the
amplitude dependence of frequency is not necessarily a fundamental property of
nonlinear dynamical phenomena \cite{Chand1}. In this case, namely, $\lambda>0$, the system admits the
explicit sinusoidal periodic solution 
\begin{eqnarray}
x(t)=\frac{A\sin{(\omega t+\delta)}}
{1-(\frac{k}{\omega})A\cos{(\omega t+\delta)}},\qquad 
0\leq A <\frac{\omega}{k},\qquad \omega=\sqrt{\lambda},
\label{lam102}
\end{eqnarray}
where $A$ and $\delta$ are arbitrary constants. The system (\ref{lam101})
exhibits certain other unusual properties also. For more details on the dynamics of
this equation one may refer to \cite{Chand1}.

In the same paper, we have also noted that the solution (\ref{lam102}) can be
derived unambiguously by introducing a nonlocal transformation,
\begin{eqnarray}
U=\displaystyle{x e^{k\int_0^t x(t')dt'}},
\label {lam103}
\end{eqnarray}
and transforming equation~(\ref{lam101}) to the linear
harmonic oscillator equation, 
\begin{eqnarray}
\ddot{U}+\lambda U=0.
\label {horm1}
\end{eqnarray}
For $\lambda>0$, the solution is
$U=A\sin(\omega t+\delta)$,
where $A$ and $\delta$ are arbitrary constants and the frequency, 
$\omega=\sqrt{\lambda}$,
is independent of the amplitude. Using (\ref{lam103}) one finds that
\begin{eqnarray}
\dot{x}=\frac{\dot{U}}{U}x-k x^2.
\label {lam104}
\end{eqnarray}
Integrating Riccati type equation~(\ref{lam104}) and absorbing the integration constant with the
existing ones,  one can obtain the solution (\ref{lam102}). We mention here
that the solution (\ref{lam102}) can be derived in a number of ways, for more
details one may again refer to \cite{Chand1} and \cite{Carinena}. One may also note
that in
the case $\lambda=0$ the transformation (\ref{lam103}) 
modifies equation (\ref{lam101}) to the free particle equation. The linearization and
integrability properties of equation (\ref{lam101}) with $\lambda=0$ has received a
considerable attention in the nonlinear dynamics literature, 
see for example, Refs. \cite{Hyet,mahomed:1985,leach:1988a,Feix}.

Now the question that naturally arises is whether the nonlocal transformation 
which connects the linear and nonlinear oscillators, namely, Eqs. (\ref{lam101})
and (\ref{horm1}) is a rare one or there exists a wider class of nonlinear
dynamical systems that are connected with linear oscillators in a hidden way?
Our studies reveal 
that there exists a class of linear oscillators that are connected with
nonlinear oscillator equations through nonlocal transformations.
For example, let us consider a damped linear harmonic oscillator, instead of the
undamped linear
 harmonic oscillator, that is,
\begin{eqnarray}            
\ddot{U}+c_1 \dot{U}+c_2 U=0,
\label {nld01}
\end{eqnarray}
where $c_1$ and $c_2$ are arbitrary parameters, and consider a more general 
form of nonlocal transformation, that is,
\begin{eqnarray}            
U=x(t)e^{\int_0^t{f(x(t'))}dt'},
\label {int02}
\end{eqnarray}
where $f(x(t))$ is an arbitrary function of $x$. Now substituting  (\ref{int02})
into (\ref{nld01}) we get a nonlinear ODE of the form
\begin{eqnarray}            
\ddot{x}+(2f+xf'+c_1)\dot{x}+(f^2+c_1f+c_2)x=0,
\label {int03}
\end{eqnarray}
where prime denotes differentiation with respect to $x$. Interestingly, one can
now see that for certain specific forms of the function $f$, the associated 
nonlinear
ODEs become the well known and well studied models in the current nonlinear
dynamics literature. To cite a few, first let us consider the case $f(x)=kx$ in 
(\ref{int02}). The associated nonlinear ODE,
\begin{eqnarray}            
\ddot{x}+(c_1+3kx)\dot{x}+k^2x^3+c_1kx^2+c_2 x=0,
\label {int04}
\end{eqnarray}
is nothing but the generalized modified Emden type equation (MEE) whose 
integrability properties have been studied in detail
in Refs. \cite{Hyet,mahomed:1985,leach:1988a,Feix,Dix:1990,Chand2,Chand3,Chand4}. 
It is one of the linearizable second order
nonlinear ODEs which admits eight parameter Lie point symmetries 
\cite{mahomed:1985}.
  
Choosing $f(x)=kx^2$ in (\ref{int02}), one gets the
generalized force-free Duffing-van der Pol oscillator (DVP) equation
\begin{eqnarray}            
\ddot{x}+(c_1+4kx^2)\dot{x}+k^2x^5+kc_1x^3+c_2x=0,
\label {int05}
\end{eqnarray}
as its nonlinear counterpart. More details on the mathematical aspects and the 
underlying dynamics of this equation can be found in 
Refs. \cite{Chand3,Chand4,Smith:1961,Sawada,Gonz:1983}. 
On the other hand fixing $f(x)=\frac{k}{x^2}$,
one gets another integrable nonlinear second order ODE of the form 
\cite{Chand3}
\begin{eqnarray}            
\ddot{x}+c_1\dot{x}+c_2x+\frac{kc_1}{x}+\frac{k^2}{x^3}=0.
\label {int06}
\end{eqnarray}
The examples illustrated above clearly demonstrate that one can systematically 
classify a class of integrable nonlinear ODEs through this way.

Motivated by these observations, in this paper, we carry out a detailed
investigation on the nonlocal connection between certain linear ODEs and their 
nonlinear counterparts of any order. To make our study a systematic one we 
start our 
investigations at the level of second order and end with $n$th order
equations. Besides constructing integrable nonlinear equations we also point out the
importance of these nonlinear equations and obtain their solutions and thereby
bring out the significance of these systems.

The plan of the paper is as follows. In Sec. 2, we make a detailed study on the
nonlocal connection between damped linear harmonic oscillator equation and
certain nonlinear oscillator equations of second order and describe a method of deriving general
solutions of the nonlinear ODEs. We extend this procedure to third and $n$th
order ODEs in Secs. 3 and 4, respectively. We present a general theory which
connects the variable coefficient linear ODEs with nonlinear ODEs and the method
of finding the general solution in Sec. 5. We present our conclusion in Sec. 6.

\section{Second order ODEs}
To begin with let us consider a linear second order ODE of the form 
(\ref{nld01}), whose general solution is 
$U=a(t)$, where $a(t)$ is a known function. Now we consider a nonlocal 
transformation of the form
\begin{eqnarray}            
U=x^ne^{\int_0^t{(\beta(t') x^m+\gamma(t'))}dt'},
\label {nld02}
\end{eqnarray}
where $n$ and $m$ are constants and $\beta(t)$ and $\gamma(t)$ are 
arbitrary functions of $t$, 
and substitute it into (\ref{nld01}) so that the latter becomes a nonlinear 
second order ODE of the form
\begin{eqnarray}            
\ddot{x}+(n-1) \frac{\dot{x}^2}{x}+\frac{\beta^2}{n} x^{2m+1}+b_1(t,x)\dot{x}
+b_2(t)x^{m+1}+b_3(t)x=0,
\label {nld03}
\end{eqnarray}
where 
\begin{eqnarray}   
b_1(t,x)=\frac{1}{n}\bigg(2n\gamma+nc_1
+(m+2n)\beta x^m\bigg),\nonumber\\
b_2(t)=\frac{1}{n}\bigg( \dot{\beta}+2\gamma \beta
+c_1\beta \bigg),
\nonumber\\
b_3(t)=\frac{1}{n}\bigg(\dot{\gamma}+\gamma^2+\gamma c_1
+c_2 \bigg).
\label {nld04}
\end{eqnarray}

Obviously the transformation (\ref{nld02}) corresponds to the special case,
$f(x)=\beta(t) x^m,\;\gamma(t)=0$, in the transformation (\ref{int02}) but with the
prefactor of the exponential function taken as $x^n$, where $n$ is arbitrary.
We have been motivated to choose this form due to the examples
illustrated in the introduction. In all these cases, one might have noted that
the function
$f(x)$ is a simple polynomial one. As a consequence, we  would like to generalize
this form and see the outcome. Such a generalization is indicated in Sec. 5. 
However, here we note that the form (\ref{nld03}) itself yields several 
important nonlinear ODEs as see below.

Now the question is how to construct the general solution for the nonlinear
equation (\ref{nld03}) from the known solution of (\ref{nld01}). This can be
done by using the identity,
\begin{eqnarray}            
\frac{\dot{U}}{U}=\frac{n\dot{x}}{x}+\beta(t)
x^m+\gamma(t),
\label {nld05}
\end{eqnarray}
derived from equation~(\ref{nld02}). Since $U=a(t)$, where $a(t)$ is a
known function, the solution of the second order linear ODE 
(\ref{nld01}), equation (\ref{nld05}) can be brought to the form
\begin{eqnarray}            
\dot{x}=\bigg(\hat{a}(t)-\gamma(t)\bigg)\frac{x}{n}
-\frac{\beta(t)}{n}x^{m+1}, 
\label {nld06}
\end{eqnarray}
where $\hat{a}=\frac{\dot{a}}{a}$. Solving equation (\ref{nld06}) 
we get the general solution for the equation 
(\ref{nld03}) in the form
\begin{eqnarray}            
\fl \qquad x(t)=\displaystyle{e^{\frac{1}{n}\int_0^t(\hat{a}(t')-\gamma(t'))dt'}}
\bigg[C+\frac{m}{n}\int_0^t\bigg(\beta(t')e^{\frac{m}{n}\int_0^{t'}
(\hat{a}(t'')-\gamma(t''))dt''}\bigg)dt'\bigg]^{\frac{-1}{m}}.
\label {nld07}
\end{eqnarray}

Note that the equation (\ref{nld07}) contains three arbitrary constants (two in
$\hat{a}(t)$, since $a(t)$ is a general solution of the second order 
equation (\ref{nld01}), and
another constant $C$). However, one can absorb one 
arbitrary
constant with the other two and rewrite the solution in such a way that it contains only two
arbitrary 
constants, as we see below.

Let us assume that the solution of (\ref{nld01}) be written 
in the form 
\begin{eqnarray}            
U(t)=a(t)=I_1e^{m_1t}+I_2e^{m_2t},
\label {nld08}
\end{eqnarray}
where $I_i$'s, $i=1,2$, are integration constants and $m_i$'s, $i=1,2$, are 
the roots of the auxiliary equation associated with the ODE (\ref{nld01}). 
From (\ref{nld08}) we get 
\begin{eqnarray}            
\hat{a}(t)=\frac{\dot{a}}{a}=\frac{\dot{U}}{U}=\frac{m_1\hat{I}_1e^{m_1t}
+m_2e^{m_2t}}
{\hat{I}_1e^{m_1t}+e^{m_2t}},\label {nld10}
\end{eqnarray} 
where $\hat{I}_1=\frac{I_1}{I_2}$. Integrating both sides of the
equation (\ref{nld10}) we get
\begin{eqnarray}            
 \int_0^t\hat{a}(t')dt'=
\log(\hat{I}_1e^{m_1t}+e^{m_2t})=\log(\frac{a(t)}{I_2}).
\label {nld11}
\end{eqnarray}
Substituting the expression (\ref{nld11}) into equation (\ref{nld07}), we get a
general solution for the equation (\ref{nld03}) in the form
\begin{eqnarray}            
\fl \qquad x(t)=\displaystyle{(\frac{a(t)}{I_2})^{\frac{1}{n}}
e^{\frac{-1}{n}\int_0^t\gamma(t')dt'}}
\nonumber\\\qquad \times 
\bigg[C+\frac{m}{n}\int_0^t\bigg(\beta(t')(\frac{a(t')}{I_2})^{\frac{m}{n}}
e^{\frac{-m}{n}\int_0^{t'}\gamma(t'')dt''}\bigg)dt'\bigg]^{\frac{-1}{m}}.
\label {int13}
\end{eqnarray}

Now the general solution (\ref{int13}) contains only two arbitrary
constants, namely, $\hat{I}_1$ and $C$. In the following, we consider certain
important sub-cases of the linear oscillator (\ref{nld01}) and discuss their
nonlinear counterparts and their physical significance.

\subsection{Case (i) $c_1=0$ and $c_2=0$}
In this case the linear equation (\ref{nld01}) is nothing but the free particle
equation. The associated nonlinear equation can be fixed from (\ref{nld03}) by
restricting $c_1=0$ and $c_2=0$. The 
general solution of the nonlinear equation turns out to be
\begin{eqnarray}            
\fl\quad
x(t)=(t+I_1)^{\frac{1}{n}}e^{\frac{-1}{n}\int_0^t\gamma(t')dt'}
\bigg[C+\frac{ m}{n}\int_0^t
\bigg(\beta(t')(t'+I_1)^{\frac{m}{n}}e^{-\frac{m}{n}\int_0^{t'}\gamma(t'')dt''}
\bigg)dt'\bigg]^{\frac{-1}{m}},
\label {solu01}
\end{eqnarray}
where $I_1$ and $C$ are arbitrary constants. 

The interesting fact is that equation (\ref{nld03}) contains a family of
important nonlinear ODEs. To mention one such example let us choose 
$n=1,\;\gamma(t)=0$ and $\beta(t)=k$ so that equation
(\ref{nld03}) now becomes
\begin{eqnarray}            
\ddot{x}+(m+2)kx^m\dot{x}+k^2x^{2m+1}=0.
\label {sfeq01}
\end{eqnarray}
The invariance, integrability properties and direct linearization through 
generalized transformation of equation (\ref{sfeq01}) can be found in 
Refs. \cite{Chand3,Chand4,Feix}. From our above results, the general solution of 
(\ref{sfeq01}) can be fixed easily from (\ref{solu01}) in the form
\begin{eqnarray}            
x(t)=(t+I_1)\bigg[C+\frac{mk}{(m+1)}(I_1+t)^{m+1}\bigg]^{\frac{-1}{m}},
\label {solu01a}
\end{eqnarray}
which exactly coincides with the results obtained through other methods 
\cite{Chand3,Chand4,Feix}.

\subsection{Case (ii) $c_1=0$ and $c_2=$ constant }
In this case, from equation (\ref{nld01}), we have the linear harmonic oscillator 
equation at our hand. The 
nonlocal transformation (\ref{nld02}) transforms the linear harmonic oscillator
equation, $\ddot{U}+c_2U=0$, to
the nonlinear ODE (\ref{nld03}) with  $c_1=0$.  The general solution of
(\ref{nld03}) turns out to be 
\begin{eqnarray}            
\fl\quad
x(t)=\bigg[\cos(\omega t+\delta)\bigg]^{\frac{1}{n}}e^{\frac{-1}{n}\int_0^t 
\gamma(t')dt'}\nonumber\\\times 
\bigg[C+\frac{ m}{n}\int_0^t \bigg(\beta(t')e^{-\frac{m}{n}
\int_0^{t'} \gamma(t'')dt''}\cos^{\frac{m}{n}}(\omega t'+\delta)\bigg)dt'
\bigg]^{\frac{-1}{m}},\label {solu02}
\end{eqnarray}
where $\omega=\sqrt{c_2}$ and $\delta$ and $C$ are arbitrary constants.

Interestingly, one can deduce certain important nonlinear ODEs from
(\ref{nld03}) with $c_1=0$. One such interesting case is  
$n=1,\;m=1,\;\gamma(t)=0$ and $\beta(t)=k$. In this case we get the equation
(\ref{lam101}). The unusual dynamical properties exhibited by this 
equation has already been pointed out in the introduction.

Now we choose $n=1,\;m=-2,\;\gamma(t)=0$ and $\beta(t)=k$ in
equation (\ref{nld03}) so that the latter becomes
\begin{eqnarray}            
\ddot{x}+c_2x+\frac{k^2}{x^{3}}=0
\label {sseq02}
\end{eqnarray}
which is another important nonlinear ODE which arises in different areas of
physics and has been studied in detail in Ref. \cite{Pinney,Lewis}. The 
solution can be derived from (\ref{solu02}) in the form
\begin{eqnarray}            
x(t)=\cos(\omega t+\delta)
\bigg[C-\frac{2k}{\omega}\tan(\omega t+\delta)\bigg]^{\frac{1}{2}},\;\;
\omega =\sqrt{c_2}.
\label {solu02b}
\end{eqnarray}
Note that the solution defines a well defined harmonic periodic oscillation
with period $T=\frac{2\pi}{\sqrt{c_2}}$ which is the same as that of the 
unperturbed simple harmonic oscillator. {\it This nonlinear oscillator again 
supports
our argument that the amplitude dependence of frequency is not necessarily a
fundamental property of nonlinear oscillations}.

On the other hand fixing $n=1,\;m=2,\;\gamma(t)=-k_2$ and $\beta(t)=k_1$ in equation
(\ref{nld03}) we get
\begin{eqnarray}            
\ddot{x}+(4k_1x^2-2k_2)\dot{x}+k_1^2x^5-2k_1k_2x^3+(c_2+k_2^2)x=0,
\label {sseq03}
\end{eqnarray}
which is nothing but the generalized force-free Duffing-van der Pol nonlinear 
oscillator equation \cite{Chand3,Smith:1961,Sawada,Gonz:1983}. The general 
solution can again be fixed from (\ref{solu02}) as
\begin{eqnarray}            
\fl \qquad x(t)=\cos(\omega t+\delta)\nonumber\\\fl \qquad\qquad\times
\bigg[Ce^{-2k_2t}+\frac{k_1}{2k_2}+\frac{k_1}{2(\omega^2+k_2^2)}
(k_2\cos2(\omega t+\delta)+\omega\sin2(\omega t+\delta))\bigg]^{-\frac{1}{2}}
\label {solu02c}
\end{eqnarray}
which exactly coincides with the result obtained through other methods
\cite{Chand3,Smith:1961,Sawada,Gonz:1983}.

Finally, let us consider the case $n=1,\;m=q,\;\gamma(t)=-k_2$ 
and $\beta(t)=k_1$ in equation (\ref{nld03}). In this case we end up with the
following nonlinear ODE
\begin{eqnarray}            
\ddot{x}+((q+2)k_1x^q-2k_2)\dot{x}+[(k_1x^q-k_2)^2+c_2]x=0.
\label {sseq04}
\end{eqnarray}
Equation (\ref{sseq04}) admits the following general solution
\begin{eqnarray}            
x(t)=\cos(\omega t+\delta)e^{k_2t}
\bigg[C+qk_1\int_0^t \bigg(e^{qk_2t'}\cos^q(\omega t'+\delta)\bigg)dt'\bigg]
^{-\frac{1}{q}}.
\label {solu02d}
\end{eqnarray}
We mention here that for even values of $q$ and positive values of
$k_1$ and $k_2$ equation (\ref{sseq04}) exhibits limit cycle oscillations
\cite{Sawada}.

\subsection{Case (iii) $c_1,\;c_2=$ constant }
In this case the nonlocal transformation connects the damped harmonic 
oscillator ODE with the nonlinear ODE (\ref{nld03}). From (\ref{int13}), the 
general solution of equation (\ref{nld03}) with $c_1$ and $c_2$ taking 
constant values turns out to be
\begin{eqnarray}            
\fl\qquad\quad
x(t)=(e^{(\frac{-c_1+\omega}{2})t}+ I_1e^{-(\frac{c_1+\omega}{2})t})^{\frac{1}{n}}
e^{\frac{-1}{n}\int_0^t \gamma(t')dt'}
\nonumber\\\qquad\times 
\bigg[C+\frac{ m}{n}\int_0^t \bigg(\beta(t')
\bigg(\frac{ I_1+e^{\omega t'}}{e^{(\frac{c_1+\omega}{2})t'}}\bigg)
^{\frac{m}{n}}e^{-\frac{m}{n}\int_0^{t'} 
\gamma(t'')dt''}\bigg)dt'\bigg]^{\frac{-1}{m}},
\label {solu03}
\end{eqnarray}
where $\omega=\sqrt{c_1^2-4c_2}$ and $I_1$ and $C$ are arbitrary constants.

In this case also one can identify several physically interesting nonlinear
oscillator equations. For example, choosing $n=1,\;m=1,\;\gamma(t)=0$ and 
$\beta(t)=k$ in 
equation (\ref{nld03}) we get an equation of the form (\ref{int04}). The general
solution can be written from (\ref{solu03}) in the form 
\begin{eqnarray} 
 x(t) =\left(\frac{2c_2(I_1+e^{\omega t})}
{C e^{(\frac{c_1+\omega}{2})t}-k(c_1-\omega)I_1
-k(c_1+\omega)e^{\omega t}}\right), 
\label{the07a}
\end{eqnarray}
where $\omega=\sqrt{c_1^2-4c_2}$. Indeed our result exactly coincides with the 
result obtained in the Refs. \cite{Chand2,Chand3}.

The choice $n=1,\;m=2,\;\gamma(t)=0$ and $\beta(t)=k$
leads us to an equation of the form (\ref{int05}).
The general solution of (\ref{int05}) can be fixed easily from (\ref{solu03}) 
in the form
\begin{eqnarray} 
\fl \qquad x(t)=\bigg(\frac{2c_1c_2(I_1+e^{\omega t})^2}
{C e^{(\frac{c_1+\omega}{2})t}-kI_1^2c_1(c_1-\omega)
-kc_1(c_1+\omega)e^{2\omega t}-8kI_1c_2 e^{\omega t}}\bigg)^{\frac{1}{2}},
\label{solu03b}
\end{eqnarray}
which indeed the general solution of the genaralized force-free DVP oscillator 
equation \cite{Chand3}.

In this section, besides the general case, we considered three different linear 
equations, namely, free particle equation, linear harmonic oscillator and 
damped linear harmonic equations and demonstrated how they are related to the well 
known nonlinear models through nonlocal transformations. In the following 
section, we extend the theory to third order ODEs and study the outcome.

\section{Third order ODEs}

Let us consider a linear third order ODE of the form
\begin{eqnarray}            
\tdot{U}+c_1 \ddot{U}+c_2 \dot{U}+c_3U=0,
\label {toe01}
\end{eqnarray}
where $c_1,\;c_2$ and $c_3$ are arbitrary constants. The nonlocal 
transformation (\ref{nld02}) transforms (\ref{toe01}) to the nonlinear 
ODE of the form
\begin{eqnarray}            
\fl \qquad
\tdot{x}+[3(n-1)\frac{\dot{x}}{x}+d_1(t,x)]\ddot{x}
+(n-1)(n-2)\frac{\dot{x}^3}{x^2}+d_2(t,x)\dot{x}^2
\nonumber\\ \qquad 
+d_3(t,x)\dot{x}+\frac{\beta^3}{n}x^{3m+1}+d_4(t)x^{2m+1}+d_5(t)x^{m+1}+d_6(t)x=0,
\label {toe02}
\end{eqnarray}
where 
\begin{eqnarray}   
\fl \quad 
d_1(t,x)=\frac{1}{n}\bigg(3n\gamma+nc_1+\beta(m+3n)x^m\bigg),
\nonumber\\\fl \quad
d_2(t,x)=\frac{1}{n}\bigg(\bigg(m(m+2n)+(n-1)(m+3n)\bigg)\beta x^{m-1}
+n(n-1)(3\gamma+c_3)x^{-1}\bigg),
\nonumber\\\fl \quad
d_3(t,x)=\frac{1}{n}\bigg(\bigg(3\gamma\beta(m+2n)+\dot{\beta}(2m+3n)
+c_1(m+2n)\beta\bigg)x^{m}
\nonumber\\\qquad\qquad
+3\beta^2(m+n)x^{2m}+n\bigg(3\dot{\gamma}+3\gamma^2+2c_1\gamma+ c_2\bigg)\bigg),
\nonumber\\
\fl \quad
d_4(t)=\frac{1}{n}\bigg(3\beta\dot{\beta}+3\gamma\beta^2
+\beta^2c_1\bigg),
\nonumber\\
\fl \quad
d_5(t)=\frac{1}{n}\bigg(\ddot{\beta}+3(\dot{\gamma}\beta+\gamma\dot{\beta})
+3\beta\gamma^2+(c_1\dot{\beta}+c_2\beta)+2c_1\gamma \beta\bigg),
\nonumber\\\fl \quad
d_6(t)=\frac{1}{n}\bigg(\ddot{\gamma}+3\gamma\dot{\gamma}+\gamma^3
+c_1(\dot{\gamma}+\gamma^2)+\gamma c_2+c_3\bigg).
\label {toe03}
\end{eqnarray}

To obtain the exact solution, $x(t)$, of equation~(\ref{toe02}) we again apply the 
results of Eqs.~(\ref{nld05})-(\ref{nld07}) to the present case.
Since we are using the same form of nonlocal transformation (\ref{nld02}) to 
transform the
nonlinear ODE (\ref{toe02}) to the linear ODE (\ref{toe01}), 
the general solution
of (\ref{toe02}) looks exactly the same form as (\ref{nld07}). The only
difference is that the general solution, in the present case, contains 
four integration constants (three of them come out from the solution of the linear ODE
(\ref{toe01}) and the remaining one 
from solving (\ref{nld06})). Now let us rewrite the
general solution in such a way that it contains only three arbitrary constants by 
absorbing the fourth one. To do so let us
follow the same procedure which we adopted in the case of second order ODEs.

Let us write the general solution of equation (\ref{toe01}) in the form
\begin{eqnarray}            
U(t)=a(t)=I_1e^{m_1t}+I_2e^{m_2t}+I_3e^{m_3t},
\label {tnld08}
\end{eqnarray}
where $I_i$'s, $i=1,2,3$, are integration constants and
$m_i$'s, $i=1,2,3$, are the roots of the auxiliary equation associated with the 
differential equation (\ref{toe01}). From (\ref{tnld08}) we get 
\begin{eqnarray}            
\hat{a}(t)=\frac{\dot{a}}{a}=\frac{\dot{U}}{U}
=\frac{m_1\hat{I}_1e^{m_1t}+m_2\hat{I}_2e^{m_2t}
+m_3e^{m_3t}}
{\hat{I}_1e^{m_1t}+\hat{I}_2e^{m_2t}+e^{m_3t}},\label {tnld10}
\end{eqnarray} 
where $\hat{I}_i=\frac{I_i}{I_3},i=1,2$. Integrating both sides of
equation (\ref{tnld10}) we get
\begin{eqnarray}            
\int_0^t\hat{a}(t')dt'=
\log(\hat{I}_1e^{m_1t}+\hat{I}_2e^{m_2t}+e^{m_3t})=\log(\frac{a(t)}{I_3}).
\label {tnld11}
\end{eqnarray}
Substituting the expression (\ref{tnld11}) into (\ref{nld07}), we get the
general solution for the equation (\ref{toe02}) in the form
\begin{eqnarray}            
\fl \qquad x(t)=\displaystyle{(\frac{a(t)}{I_3})^{\frac{1}{n}}
e^{\frac{-1}{n}\int_0^t \gamma(t')dt'}}
\bigg[C+\frac{m}{n}\int_0^t\bigg(\beta(t')(\frac{a(t')}{I_3})^{\frac{m}{n}}
e^{\frac{-m}{n}\int_0^{t'} \gamma(t'')dt''}\bigg)dt'\bigg]^{\frac{-1}{m}}.
\label {tnld13}
\end{eqnarray}
We note that the general solution (\ref{tnld13}) now contains only three 
arbitrary constants, namely, $\hat{I}_1,\;\hat{I}_2$ and $C$.

Since we have three arbitrary parameters in the
linear ODE
(\ref{toe01}), one can consider certain specific
sub-cases of physical interest. In the following, we discuss some of them.

\subsection{Case (i) $c_i=0,i=1,2,3$} 
In this case, from equation (\ref{toe01}), we have a simple linear equation 
$\frac{d^3U}{dt^3}=0$ which
can be connected to the third order nonlinear ODE
(\ref{toe02}) (with $c_1,\;c_2,\;c_3=0$) through the nonlocal transformation 
(\ref{nld02}). The general solution of (\ref{toe02}) can be fixed easily from
(\ref{tnld13}) in the form
\begin{eqnarray}            
\fl\qquad
x(t)=\bigg(\frac{1}{2}t^2+I_1t+I_2\bigg)^{\frac{1}{n}}e^{\frac{-1}{n}
\int_0^t\gamma(t')dt'}\nonumber\\\qquad\times 
\bigg[C+\frac{ m}{n}\int_0^t 
\bigg(\beta(t')\bigg(\frac{1}{2}t'^2+I_1t'+I_2\bigg)^{\frac{m}{n}}
e^{-\frac{m}{n}\int_0^{t'}\gamma(t'')dt''}
\bigg)dt'\bigg]
^{\frac{-1}{m}},\label {solu04}
\end{eqnarray}
where $I_1,I_2$ and $C$ are arbitrary constants.

Equation (\ref{toe02}) includes a class of important nonlinear systems. For
example, choosing $n=1,\;m=1,\;\gamma(t)=0$ and 
$\beta(t)=k$ equation (\ref{toe02}) becomes
\begin{eqnarray}            
\tdot{x}+4kx\ddot{x}+3k\dot{x}^2+6k^2x^2\dot{x}+k^3x^4=0.
\label {tfeq01}
\end{eqnarray} 
Equation (\ref{tfeq01}) is a special case of the Chazy equation XII
(with $N = 2$ and parametric restrictions $A = 0, B = 0$ in 
Ref. \cite{Cosgrove}), which has been studied
in detail in Refs. \cite{Chazy,Halburd,Cosgrove,Mugan,Euler,Euler:2005/06,
Chand5}. The 
general solution can be fixed easily from (\ref{solu04}) in the form
\begin{eqnarray}            
x(t)=\bigg(\frac{\frac{k t^2}{2}+I_1t+I_1I_2}{\frac{k^2t^3}{6}
+\frac{k I_1}{2}t^2+k I_1I_2t+I_1I_3}\bigg).
\label {solu04a}
\end{eqnarray}

\subsection{Case (ii) $c_3=0,\;c_2=0$ and $c_1=$ constant }
In this case, the linear equation (\ref{toe01}) becomes
$\tdot{U}+c_1 \ddot{U}=0$ and the 
nonlocal transformation (\ref{nld02}) transforms this ODE to
the form (\ref{toe02}) with  $c_2,c_3=0$.  
In this case we have the general solution of the form
\begin{eqnarray}            
\fl\quad
x(t)=\bigg(e^{-c_1t}+I_1t+I_2\bigg)^{\frac{1}{n}}
e^{\frac{-1}{n}\int_0^t\gamma(t')dt'}
\nonumber\\\times 
\bigg[C+\frac{ m}{n}\int_0^t
\bigg(\beta(t')\bigg(e^{-c_1t'}+I_1t'+I_2\bigg)^{\frac{m}{n}}
e^{-\frac{m}{n}\int_0^{t'}\gamma(t'')dt''}\bigg)dt'\bigg]^{\frac{-1}{m}},
\label {solu06}
\end{eqnarray}
where $I_1,I_2$ and $C$ are arbitrary constants.  Choosing 
$n=1,\;m=1,\;\gamma(t)=0$ and $\beta(t)=k$ equation (\ref{toe02}) becomes
\begin{eqnarray}            
\fl \qquad \qquad \tdot{x}+(c_1+4kx)\ddot{x}+3k\dot{x}^2+3k(c_1+2kx)x\dot{x}
+(c_1+kx)k^2x^3=0.\label {tfeq01a}
\end{eqnarray} 
The general solution of equation (\ref{tfeq01a}) can be fixed easily from 
(\ref{solu06}) in the form
\begin{eqnarray}            
x(t)=\bigg(\frac{e^{-c_1t}+I_1t+I_2}
{C+\frac{k}{2c_1}(2e^{-c_1t}+c_1I_1t^2+2c_1I_2t)}\bigg).
\label {solu06a}
\end{eqnarray}

\subsection{Case (iii) $c_1=0,c_2=0$ and $c_3=$ constant}
In this case, the linear ODE (\ref{toe01}) assumes the form 
$\tdot{U}+c_3 U=0$ and the 
nonlocal transformation (\ref{nld02}) transforms this equation, 
$\tdot{U}+c_3 U=0$, to
the nonlinear ODE (\ref{toe02}) with  $c_1,c_2=0$.  The general solution of
(\ref{toe02}) turns out to be
\begin{eqnarray}            
\fl\quad
x(t)=\bigg(I_1e^{-kt}+e^{\frac{k}{2}t}\cos(\frac{\sqrt{3}k}{2}t
+\delta)\bigg)^{\frac{1}{n}}e^{\frac{-1}{n}\int_0^t\gamma(t')dt'}
\nonumber\\\fl \qquad\quad\times 
\bigg[C+\frac{ m}{n}\int_0^t \bigg(\beta(t')e^{-\frac{m}{n}
\int_0^{t'}\gamma(t'')dt''}\bigg(I_1e^{-kt'}+e^{\frac{k}{2}t'}
\cos(\frac{\sqrt{3}k}{2}t'+\delta)\bigg)^{\frac{m}{n}}\bigg)dt'
\bigg]^{\frac{-1}{m}},\label {solu05}
\end{eqnarray}
where $c_3=k^3$ and $I_1,C$ and $\delta$ are arbitrary constants. By choosing
 $n=1,\;m=1,\;\gamma(t)=0$ and 
$\beta(t)=k$ in equation (\ref{toe02}) one can get a special case of the 
Chazy equation XII (with $N = 2$ and parametric restrictions $A = c_3$ 
and $B = 0$ in Ref. \cite{Cosgrove})
\begin{eqnarray}            
\tdot{x}+4kx\ddot{x}+3k\dot{x}^2+6k^2x^2\dot{x}+k^3x^4+c_3x=0.
\label {tfeq01b}
\end{eqnarray} 
The general solution of (\ref{tfeq01b}) can be fixed easily from (\ref{solu05})
in the form
\begin{eqnarray}     
x(t)=\bigg(\frac{I_1e^{-kt}+e^{\frac{k}{2}t}\cos(\frac{\sqrt{3}k}{2}t+\delta)}
{C-\frac{k}{2k} (2I_1e^{-kt}-
e^{\frac{k}{2}t}(\cos(\frac{\sqrt{3}k}{2}t+\delta)
+\sqrt{3}\sin(\frac{\sqrt{3}k}{2}t+\delta))}\bigg).
\label {solu05a}
\end{eqnarray}

\subsection{Case (iv) $c_i=$ constant, $i=1,2,3$ }
In this case the nonlocal transformation transforms the nonlinear ODE 
(\ref{toe02})
to the linear ODE (\ref{toe01}). The general solution of the latter becomes
\begin{eqnarray}            
\fl\quad
x(t)=\bigg(e^{m_1t}+I_1e^{m_2t}+I_2e^{m_3t}\bigg)^{\frac{1}{n}}
e^{\frac{-1}{n}\int_0^t\gamma(t')dt'}
\nonumber\\\fl\qquad\qquad\times 
\bigg[C+\frac{ m}{n}\int_0^t
\bigg(\beta(t')\bigg(e^{m_1t'}+I_1e^{m_2t'}+I_2e^{m_3t'}\bigg)^{\frac{m}{n}}
e^{-\frac{m}{n}\int_0^{t'}\gamma(t'')dt''}\bigg)dt'\bigg]^{\frac{-1}{m}},
\label {solu07}
\end{eqnarray}
where $I_1,I_2$ and $C$ are arbitrary constants and
$m_i^3+c_1m_i^2+c_2m_i+c_3=0,\;i=1,2,3$, and
$m_{2,3}=\frac{1}{2}[-(m_1+c_1)\pm\sqrt{-3m_1^2-2c_1m_1+c_1^2-4c_2}]$.

Equation (\ref{toe02}) includes a class of important nonlinear systems. For
$n=1,\;m=1,\;\gamma(t)=0$ and $\beta(t)=k$ equation (\ref{toe02}) becomes  
\begin{eqnarray}            
\tdot{x}+(c_1+4kx)\ddot{x}+3k\dot{x}^2
+(c_2+3kc_1+6k^2x)x\dot{x}\nonumber\\
\qquad\qquad\qquad\qquad\qquad +(c_1+kx)k^2x^3+c_2kx^2+c_3x=0.
\label {tfeq01c}
\end{eqnarray}
Equation (\ref{tfeq01c}) is a generalized version of the above special cases 
(\ref{tfeq01}) and (\ref{tfeq01b}) of the Chazy equation XII, with 
additional terms. 
The general solution of (\ref{tfeq01c}) can be fixed easily from (\ref{solu07}) 
in the form
\begin{eqnarray}            
\fl \qquad x(t)=\bigg(e^{m_1t}+I_1e^{m_2t}+I_2e^{m_3t}\bigg)
\bigg[C+k\bigg(\frac{e^{m_1t}}{m_1}+\frac{I_1}{m_2}e^{m_2t}
+\frac{I_2}{m_3}e^{m_3t}\bigg)
\bigg]^{-1}.
\label {solu07a}
\end{eqnarray}
To our knowledge, the above solution is a new one. For a detailed 
description of Chazy class of third order ordinary differential equations, 
their analytic properties and their general solutions one may refer to 
Ref. \cite{Cosgrove}, besides the original work of Chazy \cite{Chazy}.

In this section, we demonstrated the nonlocal connection that exists between
linear and nonlinear ODEs of third order. In the following section, we extend the
theory to $n$th order ODEs.

\section{$n$th order equations}
Let us consider a most general linear ODE of the form
\begin{eqnarray}            
\bigg(\frac{d^n}{dt^n}+c_1\frac{d^{(n-1)}}{dt^{(n-1)}}
+\ldots+c_{n-1}\frac{d}{dt}+c_n\bigg)U(t)=0, \label {noe01}
\end{eqnarray}
where $c_i$'s, $i=1,2,\ldots n$, are arbitrary constants. The nonlocal 
transformation (\ref{nld02}) connects (\ref{noe01}) to the nonlinear 
ODE of the form
\begin{eqnarray}            
\bigg(D_{h}^{(n)}+c_1D_{h}^{(n-1)}+\ldots+c_{n-1}D_{h}^{(1)}
+c_n\bigg)x^n=0,
\label {noe02}
\end{eqnarray}
where $D_{h}^{(n)} = 
(\frac{d}{d t}+\beta(t) x^m+\gamma(t))^n$. For $n=2$ and $3$  
equation (\ref{noe02}) coincides with (\ref{nld03}) and  (\ref{toe02}) 
respectively.
Repeating the procedure described in Secs. 2 and 3 one can derive the general
solution of the $n$th order nonlinear ODE, (\ref{noe02}), in the form
\begin{eqnarray}            
\fl \qquad x(t)=\displaystyle{e^{\frac{1}{n}\int_0^t(\hat{a}(t')-\gamma(t'))dt'}}
\bigg[C+\frac{m}{n}\int_0^t\bigg(\beta(t')e^{\frac{m}{n}\int_0^{t'}(\hat{a}(t'')
-\gamma(t''))dt''}\bigg)dt'\bigg]^{\frac{-1}{m}}.
\label {nnld07}
\end{eqnarray}
Here again $\hat{a}(t)=\frac{\dot{a}}{a}$, where $a$ is the known solution of 
equation (\ref{noe01}).
The solution contains $n+1$ integration constants, namely, $I_1,I_2,\ldots I_n$
and $C$. However, as was done in the second and third order ODEs,
respectively, we can absorb one integration constant and rewrite the solution 
in such a way
that it contains only $n$ integration constants. Doing so we arrive at
\begin{eqnarray}            
\fl \qquad x(t)=\displaystyle{(\frac{a(t)}{I_n})^{\frac{1}{n}}
e^{\frac{-1}{n}\int_0^t\gamma(t')dt'}}
\bigg[C+\frac{m}{n}\int_0^t\bigg(\beta(t')(\frac{a(t')}{I_n})^{\frac{m}{n}}
e^{\frac{-m}{n}\int_0^{t'}\gamma(t'')dt''}\bigg)dt'\bigg]^{\frac{-1}{m}}.
\label {noe09}
\end{eqnarray}

By appropriately fixing the constants and arbitrary functions in equation
(\ref{noe02}) one can deduce the nonlinear equation of interest. The general 
solution can also be fixed unambiguously from equation (\ref{noe09}).

\section{General Theory}
So far we focussed our attention only on the cases in which the nonlocal
transformation that connectes the nonlinear ODEs with constant 
coefficient linear ODEs. However, one can also explore the hidden connection
between nonlinear ODEs and variable coefficient linear ODEs. In the following we 
briefly discuss the underlying theory.

Let us begin with a $n$th order linear ODE of the
form
\begin{eqnarray} 
\bigg(\frac{d^n}{dt^n}+c_1(t)\frac{d^{(n-1)}}{dt^{(n-1)}}
+\ldots+c_{n-1}(t)\frac{d}{dt}+c_n(t)\bigg)U(t)=0,
\label {gme03}
\end{eqnarray}
where $c_i$'s, $i=1,2,\ldots n$, are 
functions of $t$, whose general solution be $U=a(t)$, where $a(t)$ is some known 
function. Now we consider a general nonlocal transformation of the form
\begin{eqnarray}            
U(t)=g(t,x)e^{\int_0^t{f(t',x)}dt'},
\label {gme01}
\end{eqnarray}
where $f(t,x)$ and $g(t,x)$ are arbitrary functions of $t$ and $x$, and
substitute it into (\ref{gme03}) so that the latter becomes a nonlinear ODE 
of the form
\begin{eqnarray}  
\bigg(D_{h}^{(n)}+c_1(t)D_{h}^{(n-1)}+\ldots
+c_{n-1}(t)D_{h}^{(1)}+c_n(t)\bigg)g(t,x)=0, 
\label {gme04}
\end{eqnarray}
where $D_{h}^{(n)} = (\frac{d}{dt}+f(t,x))^n$.
Since we have assumed that the multiplicative function $g(t,x)$ is a function 
of the 
variables $t$ and $x$, the associated nonlinear ODE, (\ref{gme04}), takes a very
complicated form. However, the task is to deduce 
the interesting cases from (\ref{gme04})
which can be integrated explicitly. Again using the identity
\begin{eqnarray}            
\frac{\dot{U}}{U}=\frac{g_x\dot{x}+g_t}{g}+f,
\label {gme02}
\end{eqnarray}
with $U=a(t)$, equation (\ref{gme02}) can be brought to the form
\begin{eqnarray}            
\dot{x}=\bigg(\hat{a}(t)-f(t,x)\bigg)\frac{g(t,x)}{g_{x}(t,x)}
-\frac{g_{t}(t,x)}{g_{x}(t,x)}, \quad \hat{a}=\frac{\dot{a}}{a}.
\label {gme08}
\end{eqnarray}
Now, it is well known that only for certain specific forms of $f$ and $g$, 
equation (\ref{gme08}) can be 
integrated \cite{Murphy:1969}. At least for these cases, the general solution for the
nonlinear ODE (\ref{gme04}) can be obtained explicity. In this way one can
classify a class of integrable nonlinear ODEs of any order through nonlocal
transformations of the form (\ref{gme01}). 

We note that the equation (\ref{gme04}) can also be obtained from equation 
(\ref{gme03}) by using a simple ``gauge transformation"  
\begin{eqnarray}            
\frac{d^n}{dt^n}&&\rightarrow D_{h}^{(n)} =\bigg(\frac{d}{d t}+f(t,x)\bigg)^n,
\quad f(t,x)=\frac{d}{dt}h(t,x)\nonumber
\end{eqnarray}
and
\begin{eqnarray} 
U(t)&&\rightarrow g(x,t). \label {gt01}\nonumber
\end{eqnarray}
Finally, we mention that in the case $f(t,x)=0$ the nonlocal transformation,
(\ref{gme01}), becomes purely a local one and the resultant linearizable
equations can be directly obtained from equation (\ref{gme04}). 

\section{Conclusion}
In this paper, we have developed a novel way of identifying integrable nonlinear
ODEs by connecting linear and nonlinear oscillator equations of any order
through suitable nonlocal transformations. The proposed method is simple and 
straightforward. More importantly, our
procedure not only identifies a class of integrable nonlinear ODEs of any order 
but also
unambiguously gives their underlying solutions and thereby leads to the complete
understanding of the dynamics of the given nonlinear system. As we have pointed 
out in the introduction, the modified Emden type
equation (\ref{lam101}) admits certain unusual nonlinear dynamical properties. 
The dynamical properties underlying other integrable nonlinear ODEs still
remain to be explored in detail and are possibly worth investigating further.
It is also of 
interest to investigate the existence of the above types of nonlocal 
connections in the case of 
partial differential equations as well, which we hope to pursue further. 
 
\section*{Acknowledgments}
The work of VKC is supported by Council of Scientific and Industrial Research 
in the form of a Senior Research Fellowship.  The work of ML forms part of the
Department of Science and Technology, Government of India, research project.

\end{document}